\documentclass[notitlepage,aps,prl,showpacs,superscriptaddress,twocolumn,amssymb,amsfonts]{revtex4-1}

\usepackage{amsfonts}
\usepackage{amssymb}
\usepackage{amsmath}
\usepackage{amsthm}
\usepackage{dsfont}
\usepackage{mathtools}
\usepackage{stackrel}
\usepackage{color}

\usepackage[pdftex]{graphicx,hyperref}

\usepackage{hyperref}

\begin{document}

\title{Obtaining highly-excited eigenstates of many-body localized Hamiltonians by the density matrix renormalization group }

\author{Vedika Khemani}
\affiliation{\mbox{Physics Department, Princeton University, Princeton, NJ 08544, USA}}
\affiliation{\mbox{Max-Planck-Institut f\"ur Physik komplexer Systeme, N\"othnitzer Str.\ 38, 01187 Dresden, Germany}}

\author{Frank Pollmann}
\affiliation{\mbox{Max-Planck-Institut f\"ur Physik komplexer Systeme, N\"othnitzer Str.\ 38, 01187 Dresden, Germany}}

\author{S. L. Sondhi}
\affiliation{\mbox{Physics Department, Princeton University, Princeton, NJ 08544, USA}}
\affiliation{\mbox{Max-Planck-Institut f\"ur Physik komplexer Systeme, N\"othnitzer Str.\ 38, 01187 Dresden, Germany}}

\begin{abstract}
The eigenstates of many-body localized (MBL) Hamiltonians exhibit low entanglement. We adapt the highly successful density-matrix renormalization group method, which is usually used to find modestly entangled ground states of local Hamiltonians, to find individual highly excited eigenstates of MBL Hamiltonians. The adaptation builds on the distinctive spatial structure of such eigenstates. We benchmark our method against the well studied random field Heisenberg model in one dimension. At moderate to large disorder, the method successfully obtains excited eigenstates with high accuracy, thereby enabling a study of MBL systems at much larger system sizes than those accessible to exact-diagonalization methods. 

\end{abstract}

\maketitle

\noindent
{\bf Introduction:} 
Many-body localization generalizes Anderson localization to interacting systems and entails disorder induced breakdown of ergodicity and thermalization. Its existence was only recently settled, following precursors \cite{Fleishman:1980,  AGKL, Gornyi:2005}, via a series of perturbative arguments \cite{Basko:2006hh, Imbrie}  and numerical studies  \cite{Oganesyan:2007ex, Pal:2010gr, Znidaric:2008cr, Bardarson:2012gc, Luitz:2015}. An intense effort has followed  revealing an extremely rich set of properties exhibited by MBL systems; see e.g. the review \cite{Nandkishore_2015} and references therein.
This work has revealed the centrality of many-body eigenstates to understanding a regime where quantum statistical mechanics simply does not apply. The phase transition between the ergodic and localized regimes\cite{Pal:2010gr,Luitz:2015,Vosk2014tr,Potter2015ab, Chandran2015cr, Grover2014,Kjaell:2015df} involves a singular change in the entanglement entropy of eigenstates from volume law to area law. Moreover, further transitions in the localized regime can involve the development of \emph{eigenstate order} \cite{Kjaell:2015df,Huse:2013bw,Pekker:2013vt,Vosk2014ept,Chandran:2013uz,Bahri:2013ux}, whereby individual highly-excited eigenstates can display patterns of order (both symmetry-breaking and topological) that may even be forbidden in an equilibrium setting. Such {\it eigenstate phase transitions} can take place at $T = \infty$ and even while usual thermodynamic functions remain non-singular. As such transitions are completely invisible to the traditional tools for studying finite energy-density phases, they necessitate a study of individual highly excited eigenstates.

Since typical MBL eigenstates have only local, area law, entanglement \cite{Pal:2010gr,Bauer:2013jw}---although deviations from the area law due to rare many-body resonances and Griffiths effects are a complication to bear in mind---the well known connection between area laws and matrix-product state (MPS)/tensor-network representations of many-body states \cite{Gottesman:2009aa, Schuch:2008ty, PhysRevLett.114.170505,Verstraete:2006fg} implies that they can be efficiently described, even at large $L$. Indeed, Pekker and Clark \cite{Pekker:2014ux} have examined the unitary operators that exactly diagonalize fully MBL (fMBL) systems and shown that they can be represented efficiently---in contrast to delocalized systems which require the full many-body Hilbert space for their specification. Parallel work \cite{Chandran:2014} argued for the existence of a single ``spectral tensor network'' that efficiently represents the entire eigenspectrum of fMBL systems. Recently the present authors and J.~I.~Cirac developed an efficient variational algorithm
 \cite{Pollmann:2015umpo} to actually {\it find} an approximate, compact representation of the diagonalizing unitary for fMBL systems---and hence obtain {\it all} the eigenstates. This algorithm captures the gross features of the spectrum very well, but does not target individual eigenstates to high accuracy. Here we describe an alternative, complementary, procedure that can be used obtain specific excited MBL eigenstates to high accuracy for large system sizes.

Our approach is directly inspired by the density matrix renormalization group (DMRG) \cite{Schollwock2011,White:1992} which has been used to great effect to obtain modestly entangled ground states in low-dimensional systems. In the MPS formalism, the DMRG algorithm variationally optimizes the MPS to minimize the ground state energy of a given Hamiltonian $H$. Naively, we
could modify this algorithm for MBL systems by targeting the eigenstate with energy closest to a specified excitation energy. However this is, in its simplest form, problematic due to the extremely small---$O(e^{-L})$---generic many-body level spacings as we will 
explicitly show below. 

We show that this problem can be overcome by making use of a defining characteristic of MBL phases, namely the existence of an emergent set of $L$ commuting $Z_2$-valued local integrals of motion (often called``l-bits") \cite{Huse:2014uy,Serbyn:2013cl,Chandran:2015df, Ros2015}.
Importantly, neighboring eigenstates with respect to the energy differ extensively in their spatial properties---we must typically flip $O(L)$ l-bits to go between them---while there is a (soft) gap to excitations with finite numbers of l-bit flips. This leads to a natural algorithm in which we select excited eigenstates based on their overlap with particular, localized spatial patterns instead of their proximity to particular energies. By this overlap metric, ``nearby'' states differ by a few [O(1)] flips of local ``l-bits. But such states are typically far separated in energy and thus the danger of mixing in eigenstates with exponentially small energy splittings is minimized. 

We start with a brief review of the ground state DMRG method before describing our modified \emph{DMRG-X} procedure.  We then apply the method to the random field Heisenberg chain and evaluate our results using various metrics like energy variances and overlaps with exact eigenstates. For strong enough disorder, we obtain eigenstates with machine-precision variance and find a rapid convergence of variances with bond dimension. Finally, we use our eigenstates to efficiently compute local-expectation values and demonstrate the failure of the eigenstate thermalization hypothesis (ETH) \cite{Deutsch:1991ju,Srednicki:1994dl,Rigol:2008bf} in the MBL phase. We note unpublished work \cite{ PekkerTalk2014, YuPekkerClark} that also generalizes DMRG to highly excited states using a more complex energy based targeting approach. 

\noindent
{\bf DMRG-X Method:}
The proposed method is a reformulation of the standard DMRG algorithm \cite{White:1992,Schollwock2011} to find highly-excited states of MBL systems. For a one-dimensional system of $L$ sites, a general quantum state $|\Psi\rangle$ can  be written in the following MPS form:
\begin{equation}
	|\Psi \rangle = \sum_{j_1, \ldots, j_L} B^{[1]j_1}B^{[2]j_2} \ldots B^{[L]j_L} | j_1, \ldots ,j_{L} \rangle.  \label{eq:mps}
\end{equation}
Here,  $B^{[n]j_n}$ is a $\chi_{n} \times \chi_{n+1}$ matrix and $|j_n\rangle$ with $j_n=1,\dots,d$ is a basis of local states at site $n$ (for a spin 1/2 system, $d = 2$). Each matrix product $\prod_i B^{[i]j_i}$ in~Eq.~(\ref{eq:mps}) produces a complex number which is the amplitude of $|\Psi \rangle$ on the basis state $|j_1 \cdots j_L \rangle$. The key insight behind the success of DMRG is that ground states of one dimensional  systems are efficiently approximated by MPS \cite{Verstraete:2006fg}. 
Starting from an initial random MPS, the DMRG algorithm sweeps through the system and iteratively optimizes the matrices $B^{[n]j_n}$ by locally minimizing the energy with respect to a given Hamiltonian $H$ \footnote{See Supplementary Material for details, which includes Refs\cite{VidalCanonical, KjallSpin2}.}. 
For the commonly used two-site update which simultaneously updates the matrices $B^{[n]j_n}$ and $B^{[{n+1}]j_{n+1}}$, an effective Hamiltonian $\mathcal{H}$ is constructed by projecting $H$ to a mixed $\chi_{n}\chi_{n+2}d^2$ dimensional basis . Here, the local basis states $|j_n\rangle|j_{n+1}\rangle$ represent the two updated sites, and the eigenstates of the reduced density matrix $|\chi_{n}\rangle_L$ and $|\chi_{n+2}\rangle_R$ compactly represent the environment to the left and right of the updated sites. The ground state of $\mathcal{H}$ is found and the matrices on sites $n, n+1$ are updated. The procedure is then repeated for all sites until convergence is achieved.

The DMRG-X method for finding excited eigenstates proceeds similarly to the standard DMRG algorithm in that we iteratively optimize an MPS.
The key difference is that the algorithm does not attempt convergence in the energy of $\mathcal{H}$ but instead in  
the local spatial structure of the eigenstate.
We start by initializing the algorithm with a product state that has a finite overlap with some l-bit state, e.g., for the random-field Heisenberg model discussed below, we choose random states in the $\sigma^z$  basis of the form  $|\psi\rangle_0 = | \uparrow \downarrow \downarrow \uparrow ...\downarrow\uparrow\rangle$.  We start our DMRG-X algorithm with the following local two-site update :
 (i) Construct the effective Hamiltonian $\mathcal{H}$.
 (ii) Pick the eigenstate of $\mathcal{H}$ that has \emph{maximum overlap} with the current MPS.
 (iii) Update the tensors $B^{[n]j_n}$ and $B^{[{n+1}]j_{n+1}}$. To produce the data below, we use a  full diagonalization of $\mathcal{H}$ which scales as $\chi^6$ with $\chi$ being the mean bond dimension.
Alternatively, it is also possible to find a small set of $k$  eigenstates of  $\mathcal{H}$ near the energy of the current MPS and then pick the eigenstate with largest overlap. This yields an algorithm that scales approximately as $\chi^3$, but an optimal $k$ has to be found for each case. 
 
This DMRG-X prescription ensures that no individual update step of the MPS matrices results in a large spatial reorganization, which is appropriate for a localized phase.
By contrast, if we pick excited eigenstates of $\mathcal{H}$ that are closest in energy to some target energy, the exponentially small energy gaps mean that we could be picking very different eigenstates (as labeled by their l-bit quantum numbers) at each step. 
This will, in general, result in a  slow convergence and/or a final state that is a superposition of many nearby eigenstates. 

\begin{figure}[tb]
  \includegraphics[width=1.\columnwidth]{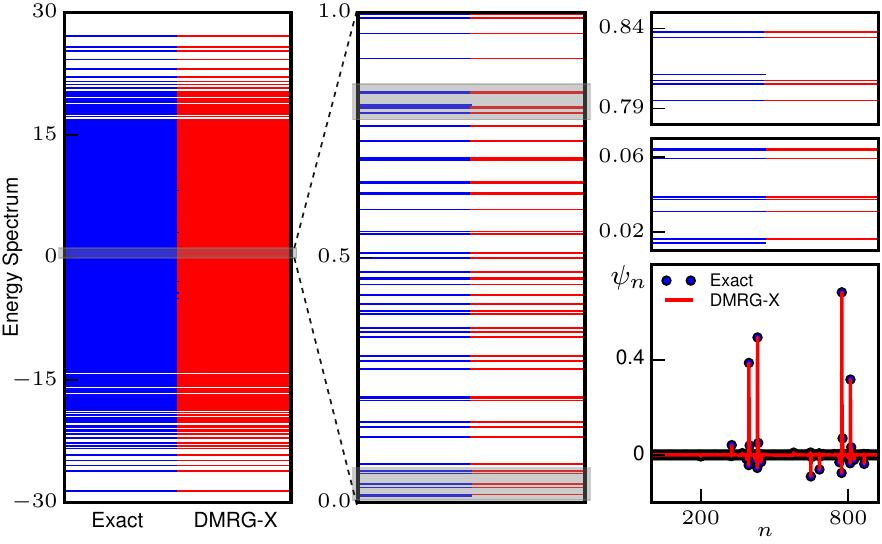} \\
  \caption{\label{fig:EDSpec} Comparison between eigenenergies obtained using exact diagonalization (blue) and variational DMRG-X (red) for a system of size $L = 12$, disorder strength $W = 8$ and bond dimension $\chi = 16$. The successive panels which zoom into the shaded regions of the spectrum show that all individual eigenenergies are obtained extremely accurately.  The bottom right panel shows the exact and variational amplitudes for a particular eigenstate with Mott-resonances, showing that the method successively captures resonant states.   }
 \end{figure}
\noindent
{\bf Comparison with ED for small systems:}  We now benchmark our method against the Heisenberg model with random $z$-directed magnetic fields:
\begin{equation}
H = J \sum_n  \vec{S}_n\cdot\vec{S}_{n+1} - \sum_n h_n S_n^z. \label{eq:ham}
\end{equation}
where $J = 1$, $\vec{S_n}$ are spin 1/2 operators and the fields $h_n$ are drawn randomly from the interval $[-W,W]$. 
This model has been studied extensively in the context of MBL and several numerical studies \cite{Pal:2010gr, Luitz:2015,Devakul:2015fg} indicate  that $H$ is  fMBL for $W \gtrsim 3.5-4$. At strong disorder, typical eigenstates look like product states in the $\sigma^z$ basis with small fluctuations. Equivalently, the ``l-bits'' $\tau_i^z$ look like $\sigma_i^z$ with exponentially decaying corrections from operators away from site $i$. As the disorder is lowered, the probability of many-body ``Mott-type'' resonances\cite{sarangetal, Mott1968}, wherein the eigenstates are approximately equal-weight superpositions of a few basis states, increases. These resonant states have energy splittings that decay exponentially with the maximum distance involved in the resonance. At even smaller $W$s, the  approach to the ergodic transition is marked by a ``Griffiths'' region \cite{sarangetal, Vosk2014tr} in which locally ergodic/critical inclusions start to proliferate.

Figure~\ref{fig:EDSpec} shows a comparison between eigenenergies obtained using exact diagonalization (blue) and variational DMRG-X (red) for a system of size $L = 12$, disorder strength $W = 8$ and bond dimension $\chi = 16$. To obtain the full spectrum, we feed the algorithm all possible $\sigma^z$ product states as initial states. We find that the variance in the energy of all variationally obtained DMRG-X eigenstates is less than machine precision ($\sim  10^{-12}$), and the overlap of these states with the exact eigenstates is unity up to machine precision.  

The zoomed in energy levels show that the method successively resolves the exponentially small splittings in the spectrum extremely accurately. However, a few exact eigenenergies have no DMRG-X partner---when two or more eigenstates of $H$ have maximum weight on the same input basis state, the input state converges to one of these eigenstates leaving the other unpaired. We can avoid this duplication by requiring every new state to be orthogonal to the prior ones, but this will not be necessary in larger systems where our goal will never be to obtain every eigenstate. 

One might worry that this is method biased towards product states and fails to capture resonant eigenstates. The bottom-right panel of Fig.~\ref{fig:EDSpec} shows a representative eigenstate with a many-body ``Mott" resonance involving a few distant basis states which is exactly captured by the variational state.  We emphasize that the algorithm only uses a product state as an initial input; after that, the algorithm converges to the previously chosen eigenstate of $\mathcal{H}$. As long as the bond dimension $\chi$ is sufficiently large for the eigenstates of $\mathcal{H}$ to capture resonances, it is easy for the algorithm to converge to a resonant superposition starting from one of the product states with significant weight in the resonant eigenstate. 

\begin{figure}[tb]
  \includegraphics[width=1.\columnwidth]{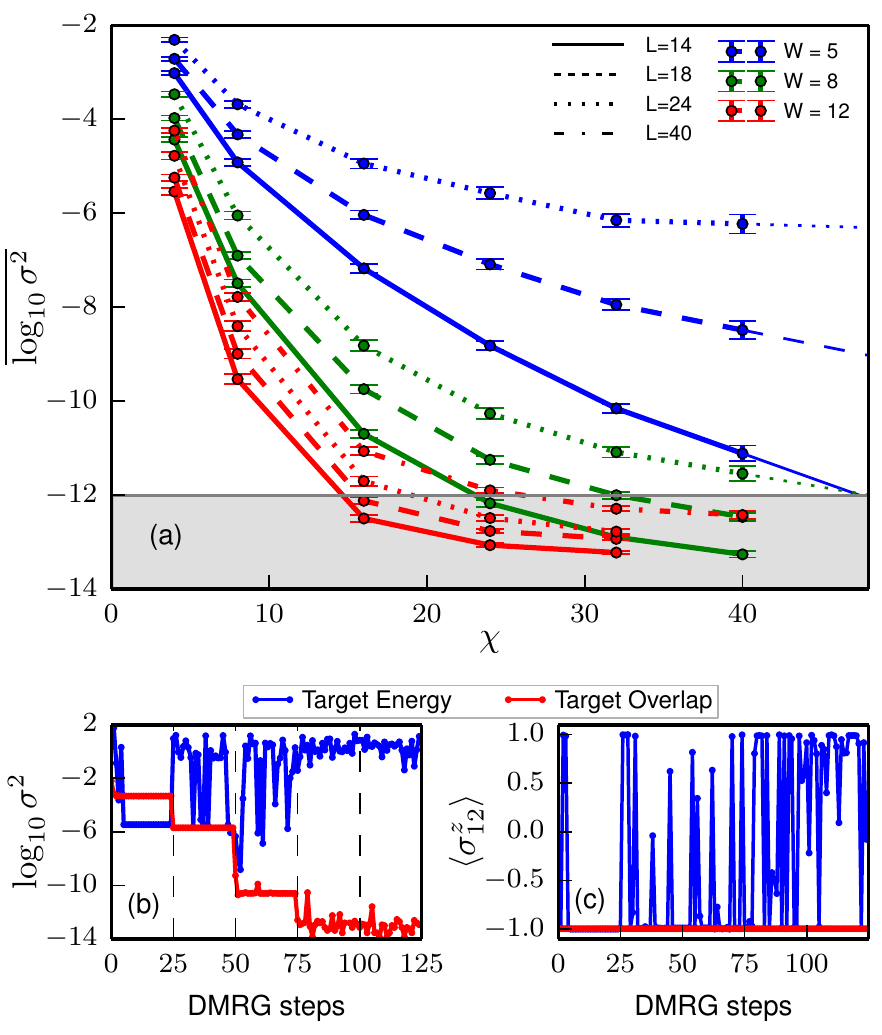} \\
  \caption{\label{fig:Variance} (a) Disorder averaged logarithm of the energy variance $\sigma^2$ plotted against bond-dimension $\chi$ for different disorder strengths $W$ and system sizes $L$. We see a rapid decrease of the typical $\sigma^2$ with $\chi$ for moderate-strong disorder, with variances falling below machine precision (shaded grey region below $10^{-12}$) at small $\chi \ll 2^{L/2}$. (b), (c) Variance and $\langle \sigma^z_{12} \rangle$ plotted against DMRG time-steps for a typical run in which $\chi$ is successively increased after each 25 steps to obtain a single eigenstate with $L = 24, W = 12$ using our overlap-based DMRG-X method (red), and a more naive energy-targeting method (blue) showing vastly better convergence for the overlap method. }
 \end{figure}
 
\noindent
{\bf Larger systems:} We now turn to an evaluation of the algorithm for system sizes inaccessible to ED by examining typical variances in the energy, $\sigma^2 = \langle H^2\rangle - \langle H \rangle^2$ for approximate eigenstates of the Hamiltonian (\ref{eq:ham}) obtained using DMRG-X at different disorder strengths $W$ and system sizes $L$. Figure \ref{fig:Variance}(a) shows the disorder averaged value of $\log_{10} \sigma^2$ as a function of bond-dimension $\chi$ for randomly chosen excited states from 200-1000 disorder samples at different values of $W$ and $L$. The grey line at $10^{-12}$  marks the approximate value of machine-precision and we average  $\log \sigma^2$ to capture typical behavior instead of deviations due to rare eigenstates. 

At strong and moderate disorder ($W = 8, 12$), we see an initial rapid decrease of $\sigma^2$ with $\chi$ followed by a saturation---the saturation is expected to happen when the bond dimension becomes large enough to capture entanglement over a correlation length $\log_2 \chi \sim \xi$.  Even at moderate disorder $W=8$, the bond dimension saturates quickly and $\chi \sim 40 \ll 2^{L/2}$ is already sufficient to capture states to machine precision accuracy! In this regime, this method can be used to really push the boundaries on the system sizes that we have been able to study through ED. As the transition to the ergodic phase is approached, locally thermal Griffiths regions become more probable and the eigenstates become more entangled. We see that the accuracy of the method for the small bond dimensions considered starts to break down around $W = 5$, though a rough extrapolation suggests that we can still make significant improvements by using larger $\chi$. The increase in variance with system-size at fixed $\chi$ is to be expected since even clean ground state DMRG methods make a constant error per unit length and yield a variance that grows with system size.

At even larger sizes or/and small $W$, inevitable locally thermal/ critical Griffiths inclusions will require special handling as a subset of the l-bits now look more delocalized and the eigenstates have a very different structure from product states within the inclusions. A comparison with ED (not shown) on a system with an artificially engineered thermal inclusion shows that the variationally obtained states correctly capture local observables away from the inclusion, but make superpositions between eigenstates that differ primarily in the inclusion region. In principle, it is possible to purify these states to obtain an eigenstate by using a hybrid energy-overlap method; an inclusion of length $\ell$ occurs with probability\cite{sarangetal} $p_W^ {\ell}$ with $p_W < 1$ and has a level spacing $\Delta \sim 2^{-\ell} $. We identify the Griffiths inclusion by looking for a diminished value of the frozen moment $|\langle \sigma_i^z \rangle|$ in the states obtained by DMRG-X. We then feed these states into a hybrid algorithm which picks states at a chosen energy from the subset of states which have large overlap with the starting state away from the inclusion.  This ensures that we're only trying to resolve the larger level spacing $\Delta \ll 2^{-L}$, while also maintaining the integrity of the state away from the inclusion.

Note that for a \emph{typical} cut somewhere along the chain, the entanglement entropy scales an an area law with co-efficient $\xi$
proportional to the localization length since the state looks thermal on length scales shorter than $\xi$ \cite{Grover2014}. This implies that the typical bond dimension scales exponentially with $\xi$. On the other hand, the \emph{maximum} entanglement entropy across all cuts in the chain scales logarithmically with $L$ (a thermal region of size $\ell$ is exponentially rare in $\ell$, but has $O(L)$ chances for occuring \emph{somewhere} in a system of size $L$, thereby giving $\ell \sim \log(L)$). This implies a polynomial scaling with $L$ for the maximum bond-dimension $\chi_{\rm max}$. Combining this with the $\chi^6 L$ scaling of the cost of the DMRG-X algorithm means that the algorithm scales exponentially in $\xi$ and linearly in $L$ if the maximum $\chi$ is fixed at some O(1) number $\sim e^{\xi}$. On the other hand, if the bond dimension is allowed to grow to achieve a certain accuracy, then the cost scales polynomially in $L$ with a power larger than 1 and dependent on $W$. 

We end with two comments. First, for large system sizes, we can randomly sample from the spectrum and approximate the underlying density of states by randomly choosing intial product states. Even though the DMRG-X sweep does not use energy targeting, we can still effectively target different energy densities. Deep in the disordered phase, the initial product state is exponentially close to an actual eigenstate, and thus $\langle H \rangle$ is almost constant during a run. 

\begin{figure}[tb]
  \includegraphics[width=1.\columnwidth]{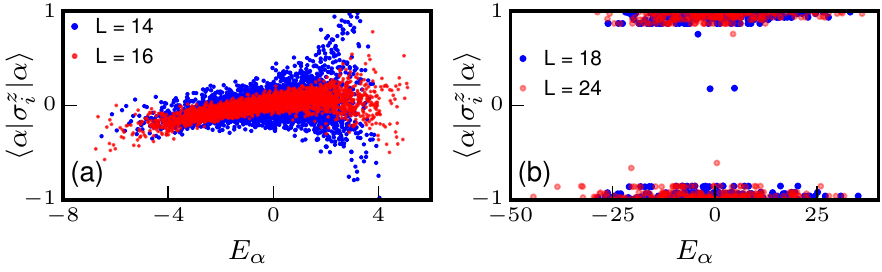} \\
  \caption{\label{fig:ETH} (a) $\langle \alpha|\sigma^z_i| \alpha \rangle$ plotted against  $E_\alpha = \langle \alpha|H| \alpha \rangle$ for an ergodic system with $W = 1.5$ and eigenstates $|\alpha \rangle$ obtained via ED. This is an ETH obeying phase where the observable varies smoothly with energy and the fluctuations decrease with increasing $L$. (b) Same quantity evaluated with $\sim 500$ variationally obtained eigenstates $|\alpha \rangle$ of an MBL Hamiltonian with $W = 8$ and $ L = 18,24$ showing a clear violation of ETH}.
 \end{figure}
 
Second, since the variationally obtained states are MPSs, few-point observables can be computed extremely efficiently. In Fig~\ref{fig:ETH}(b) we show $\langle \alpha|\sigma^z_i| \alpha \rangle$ plotted against  $E_\alpha = \langle \alpha|H| \alpha \rangle$ for $\sim 500$ variationally obtained eigenstates $|\alpha \rangle$ of an MBL Hamiltonian with $W = 8$ and $ L = 18,24$. We see a clear violation of ETH since the local observable does not vary smoothly with $E_\alpha$ and the fluctuations do not decrease with $L$. This also lends additional support that our method is correctly capturing MBL eigenstates since the violation of ETH would have been much weaker if the states $|\alpha \rangle$ were superpositions of actual eigenstates. Such a test is especially useful at large $L$s where the average level spacing is smaller than machine precision and we need to rely on methods other than the variance to diagnose the goodness of the variational states. 
Figure~\ref{fig:ETH}(a) shows the analogous calculation in an ergodic system with $W = 1.5$ and eigenstates obtained via ED. Here we do see a smooth variation of the observable with $E_\alpha$, and the characteristic decrease in fluctuations\cite{Beugeling2014ab} with increasing $L$. 

\noindent
{\bf Energy targeting:}
We now compare the convergence of our overlap method with the simplest energy targeting method which picks the eigenstate of $\mathcal{H}$ closest to a chosen energy. Figure~\ref{fig:Variance}(b) shows a typical DMRG-X run with $L=24$ and $W=12$ to obtain a single eigenstate. The bond dimension is increased every 25 steps and takes the values $\chi = (4, 8,16,24,32)$. We see that the overlap method (shown in red) converges extremely quickly each time the bond-dimension is increased and rapidly reaches machine precision. On the other hand, the energy targeting method (shown in blue) run for the same disorder realization and a target energy equal to the energy of the state obtained via the overlap method (upto 4 digits of precision) shows an extremely poor convergence and very large variances. In Fig. \ref{fig:Variance}(c) we plot the expectation value of $\sigma^z$  for a site in the middle of the chain evaluated using the states at each DMRG step. As expected, the overlap method shows very little fluctuation in this quantity, while the naive energy approach is clearly seen to be rattling between states with extremely different local quantum numbers.

\noindent
{\bf Summary and Outlook:} 
In summary, we have developed a DMRG-X method that successfully obtains \emph{highly excited} eigenstates of MBL systems to machine precision accuracy at moderate-large disorder in a time that scales only polynomially with $L$. 
This method explicitly takes advantage of the local spatial structure and order characterizing MBL eigenstates, thereby moving away from traditional energy based DMRG algorithms. 

A natural next step is to use the DMRG-X method to obtain phase boundaries between localized phases with different kinds of eigenstate order present. The nature of the phase transition between different localized phases is an important open question, and refining this technique to access these transitions at  larger system sizes should  help settle some of these questions.

\noindent
{\bf Acknowledgements:} We are indebted to David Huse for regular enlightenment, Ignacio Cirac for collaboration on closely related work and to Bryan Clark for generously sharing his unpublished work with us. This work was supported by NSF Grant No. 1311781 and the John Templeton Foundation (VK and SLS) and the Alexander von Humboldt Foundation and the German Science Foundation (DFG) via the Gottfried Wilhelm Leibniz Prize Programme at MPI-PKS (SLS) and the SFB 1143.

\bibliography{mbl}

\end{document}


\title{Supplement: Obtaining highly-excited eigenstates of many-body localized Hamiltonians by the density matrix renormalization group}

\author{Vedika Khemani}
\affiliation{\mbox{Department of Physics, Princeton University, Princeton, NJ 08544, USA}}
\affiliation{\mbox{Max-Planck-Institut f\"ur Physik komplexer Systeme, N\"othnitzer Str.\ 38, 01187 Dresden, Germany}}

\author{Frank Pollmann}
\affiliation{\mbox{Max-Planck-Institut f\"ur Physik komplexer Systeme, N\"othnitzer Str.\ 38, 01187 Dresden, Germany}}

\author{S. L. Sondhi}
\affiliation{\mbox{Department of Physics, Princeton University, Princeton, NJ 08544, USA}}
\affiliation{\mbox{Max-Planck-Institut f\"ur Physik komplexer Systeme, N\"othnitzer Str.\ 38, 01187 Dresden, Germany}}
\maketitle
\section{Method}

In this section, we briefly recapitulate the standard DMRG algorithm\cite{White:1992} implemented in the language of matrix-product states (MPSs) \cite{Schollwock2011}, before providing more details on the DMRG-X method. Section IV in Ref.~\onlinecite{KjallSpin2} also provides a very clear exposition of these numerical methods and some of our discussion closely follows this reference.

\begin{figure}[]
\includegraphics[width=0.50\columnwidth]{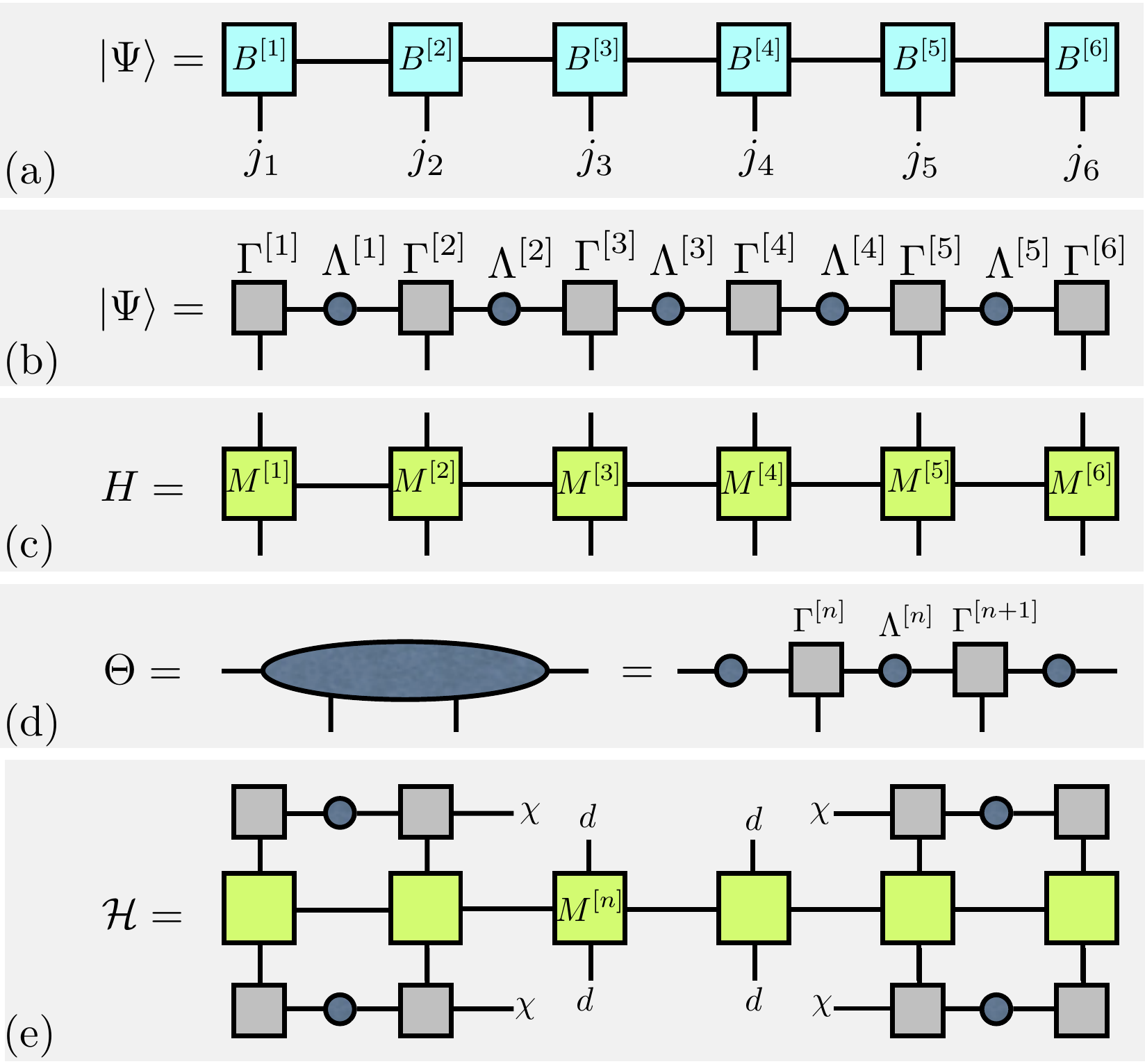} \\
 \caption{Diagrammatic representation of (a) the state $|\Psi\rangle$ as an MPS, (b) $|\Psi\rangle$ as a canonical MPS, (c) the Hamiltonian $H$ as an MPO, (d) co-efficients $\Theta$ of $|\Psi\rangle$ in the variational basis and (e) the effective Hamiltonian $\mathcal{H}$ in the variational basis.  }
 \label{fig:MPS}
\end{figure}

A general quantum state $|\Psi\rangle$ for a one-dimensional system of $L$ sites can be written in the following matrix-product state (MPS) form:
\begin{equation}
	|\Psi \rangle = \sum_{j_1, \ldots, j_L} \sum_{0< \gamma_n <= \chi_n} B^{[1]j_1}_{\gamma_1} B^{[2]j_2}_{\gamma_1 \gamma_2} \ldots B^{[L]j_L}_{\gamma_{L-1}} | j_1, \ldots ,j_{L} \rangle.  \label{eq:mps}
\end{equation}
where $|j_n\rangle$ with $j_n=1,\dots,d$ is a basis of local states at site $n$ (for a spin 1/2 system, $d = 2$ and $|j_n\rangle = |\uparrow \rangle, |\downarrow\rangle$), and the  $B^{[n]}$ are rank three tensors (except on the first and last sites where they are rank two tensors). Figure \ref{fig:MPS}(a) shows a pictorial representation of an MPS. The enternal legs $j_n$ are the ``physical'' spin indices whereas the internal legs $\gamma_n$ are the virtual indices that are contracted. Each $B^{[n]j_n}$   is a $\chi_{n} \times \chi_{n+1}$ matrix (at the boundaries $\chi_1 = \chi_{L+1} = 1$) and each matrix product $\prod_i B^{[i]j_i}$ in~Eq.~(\ref{eq:mps}) produces a complex number which is the amplitude of $|\Psi \rangle$ on the basis state $|j_1 \cdots j_L \rangle$. 

The maximum dimension $\chi$ of the $\{B^{[n]}\}$ matrices is called the bond-dimension of the MPS and low entanglement states can be efficiently represented my MPSs of bond dimension $\chi \ll d^{L/2}$. The relationship between $\chi$ and the entanglement can be made more precise by considering the Schmidt decomposition of the state $|\Psi\rangle$. For a given bipartition of the system into left and right halves, a singular value decomposition can be used to rewrite $|\Psi\rangle$ as  
$$
|\Psi\rangle = \sum_{\alpha} \Lambda_\alpha |\alpha \rangle_L |\alpha\rangle_R
$$
where the $|\alpha \rangle_{L/R}$ form orthonormal bases for the left and right halves respectively, ${}_L\langle \alpha | \beta \rangle_L = {}_R\langle \alpha | \beta \rangle_R =\delta_{\alpha \beta}$,  and the entanglement entropy of the bipartition is defined through the Schmidt values $\Lambda_\alpha$ as $S_E = -\sum_\alpha |\Lambda_\alpha|^2 \ln |\Lambda_\alpha|^2$. Following a prescription by Vidal\cite{VidalCanonical}, it is possible to define a \emph{canonical form} (Fig.\ref{fig:MPS}(b)) for the MPS by rewriting each matrix $B^{[n]j_n}$ as a product of a $\chi_{n} \times \chi_{n+1}$ dimensional complex matrix $\Gamma^{[n]j_n}$ and a square diagonal matrix $\Lambda^{[n]}$ such that matrices $\Lambda^{[n]}$ matrices contain the non-zero Schmidt values for a bipartition between sites $n$ and $n+1$
\begin{equation}
	|\Psi \rangle = \sum_{j_1, \ldots, j_L}  \Gamma^{[1]j_1} \Lambda^{[1]} \Gamma^{[2]j_2} \Lambda^{[2]} \ldots \Lambda^{[L-1]}\Gamma^{[L]j_L} | j_1, \ldots ,j_{L} \rangle = \sum_{\alpha = 1}^{\chi_{n+1}}\Lambda^{[n]}_{\alpha \alpha} |\alpha_n\rangle_L |\alpha_n \rangle_R,  
	\label{eq:canonical}
\end{equation}
and the states 
\begin{align}
|\alpha_n\rangle_L &\equiv \sum_{j_1 \cdots j_{n} } ( \Gamma^{[1]j_1} \Lambda^{[1]} \cdots \Lambda^{[n-1]}\Gamma^{n})_\alpha |j_1 \cdots j_n  \rangle \nonumber \\
|\alpha_n\rangle_R &\equiv \sum_{j_{n+1} \cdots j_{L} } ( \Gamma^{[n+1]j_{n_1}} \Lambda^{[n+1]} \cdots \Lambda^{[L-1]}\Gamma^{L})_\alpha |j_{n+1} \cdots j_L  \rangle 
\label{eq:schmidtCan}
\end{align}
 define the orthonormal Schmidt states for the left and right halves of the bipartition respectively. 
This canonical form clearly relates the bond dimension of the MPS $\chi$ to the number of Schmidt values contributing significantly to the entanglement entropy.

A standard two-site DMRG algorithm tries to find the ground state $|\psi_0\rangle$ by variationally optimizing the MPS matrices on neighboring sites to minimize the energy $\langle \psi_0| H| \psi	_0\rangle$ while keeping the rest of the chain fixed. We define a matrix-product operator (MPO) representation of $H$ exactly analagous to Eq.~\ref{eq:mps} but now using 4-index tensors $M$ as shown in Fig.~\ref{fig:MPS}(c), and then implement the following steps: 
\begin{itemize}
\item To update the MPS between sites $n$ and $n+1$, rewrite the state $|\Psi\rangle	$ in the basis spanned by the states $|j_n\rangle$, $|k_{n+1}\rangle$ and the left and right Schmidt states $|\alpha_{n-1}\rangle_L$ and $|\beta_{n+1}\rangle_R$
$$
|\psi\rangle = \sum_{j,k,\alpha,\beta}  \Theta_{\alpha \beta}^{j k} |\alpha_{n-1}\rangle_L |j_n\rangle |k_{n+1}\rangle |\beta_{n+1}\rangle
$$
where the definition of $\Theta$ is shown pictorially in Fig.~\ref{fig:MPS}(d) and follows directly from the definitions \eqref{eq:schmidtCan}. 

\item Define an effective Hamiltonian $\mathcal{H}$ as the Hamiltonian projected to the $|\alpha j k \beta\rangle$ basis. This is a $d^2 \chi^2 \times d^2 \chi^2 $ dimensional operator, again best seen pictorially in Fig.~\ref{fig:MPS}(e). 

\item Find the ground state of $\mathcal{H}$, denoted by $\tilde{\Theta}_{\alpha \beta}^{jk}$. This is the optimal state for minimizing $\langle \psi_0| H| \psi	_0\rangle$  in this subspace. 

\item Do an an SVD on $\tilde{\Theta}$ to put the MPS back in canonical form with the matrices $\Gamma^{[n]}, \Lambda^{[n]}, \Gamma^{[n+1]}$ updated. 

\item Repeat for the next pair of sites and iteratively sweep through the chain till the state converges.   

The only difference between the ground-state DMRG algorithm outlined above and DMRG-X is in step 3. In the DMRG-X algorithm, we find \emph{all} the $d^2 \chi^2$ eigenstates of $\mathcal{H}$ instead of just its ground state. We then pick $\tilde{\Theta}$ as the eigenstate of $\mathcal{H}$ with the maximum overlap with the previously found state in the iterative scheme. The algorithm is initialized with an appropriate initial state which is perturbatively ``close'' to the true eigenstates of the MBL Hamiltonian (such as a product state in the $\sigma^z$ basis).

\end{itemize}

\bibliography{mbl}